\documentclass[aps,prb,twocolumn,superscriptaddress]{revtex4}
\usepackage{graphicx}
\usepackage{latexsym}
\usepackage{amssymb}
\usepackage{amsmath}
\usepackage{amsfonts}
\usepackage{bm}
\usepackage{multirow}
\usepackage{color}

\newcommand{\beq}{\begin{equation}}
\newcommand{\eeq}{\end{equation}}
\newcommand{\beqn}{\begin{eqnarray}}
\newcommand{\eeqn}{\end{eqnarray}}
\usepackage{amsmath,amssymb,amsfonts,graphics,multirow}

\begin{document}

\title{Stable Gapless Bose Liquid Phases without Any Symmetry}

\date{\today}

\author{Alex Rasmussen}

\author{Yi-Zhuang You}

\author{Cenke Xu}

\affiliation{Department of physics, University of California,
Santa Barbara, CA 93106, USA}

\begin{abstract}

It is well-known that a stable algebraic spin liquid state (or
equivalently an algebraic Bose liquid (ABL) state) with emergent
gapless photon excitations can exist in quantum spin ice
systems~\cite{PhysRevB.69.064404,wen2003,spinice1,spinice6,slbalents,spinice2,spinice3,spinice4,spinice5},
or in a quantum dimer model on a bipartite $3d$
lattice~\cite{sondhiphoton}. This photon phase is stable against
any weak perturbation without assuming any symmetry. Further works
concluded that certain lattice models give rise to more exotic
stable algebraic Bose liquid phases with graviton-like
excitations~\cite{2006cond.mat..2443X,guwen1,PhysRevB.74.224433,guwen2,PhysRevD.81.104033}.
In this paper we will show how these algebraic Bose liquid states
can be generalized to stable phases with even more exotic types of
gapless excitations and then argue that these new phases are
stable against weak perturbations. We also explicitly show that
these theories have an (algebraic) topological ground state
degeneracy on a torus, and construct the corresponding topological
invariants.

\end{abstract}

\maketitle

\section{Introduction}

It is now well-known that quantum disordered states of many-body
systems can be fundamentally different from classical disordered
states. Without assuming any symmetry, there is simply one type of
trivial classical state, but there can be many stable quantum
disordered states. Many of these nontrivial quantum disordered
phases have a gapped spectrum and topological degeneracy on a
manifold with nontrivial
topology~\cite{wentopo1,wentopo2,wentopo3}, such as fractional
quantum Hall states. In this paper we consider another kind of
stable quantum disordered phases without assuming any symmetry.
These states are characterized by their bulk gapless bosonic modes
that {\it cannot} be interpreted as Goldstone modes. Furthermore,
physical quantities have power-law (or algebraic) correlations
instead of short-range correlations found in gapped systems.


Although such gapless states are not rare at all in condensed
matter systems, they usually occur at quantum critical points and
are protected by certain symmetries. Generically, we would expect
there to be relevant perturbations that will open the gap in these
critical states. But the examples we will discuss in this paper
all have very stable gapless bosonic modes, which are invulnerable
to any weak perturbations.
Thus, to establish that an algebraic Bose liquid (ABL) phase is
stable, we must show that \textit{all} potential gap-opening
perturbations are irrelevant at the IR fixed point of the ABL
phase. Drawing intuition from the $(2+1)d$ compact lattice U(1)
gauge field, we must demonstrate not only a direct mass term of
these gapless modes are forbidden, but also that the space-time
topological defects in the dual picture must also be suppressed
(or irrelevant).



A few examples of this type of states are already known. In
Ref~\onlinecite{PhysRevB.69.064404,wen2003,sondhiphoton} a stable
ABL phase with photon like excitations were proposed, and it has
attracted great interests~\cite{spinice1,spinice6,slbalents}. So
far compelling experimental evidences for such liquid states have
been found~\cite{spinice2,spinice3,spinice4,spinice5}. Later, a
different type of ABL phase with graviton like excitation was
studied in
Ref.~\onlinecite{2006cond.mat..2443X,guwen1,PhysRevB.74.224433}.
It turns out that the graviton-ABL state has a close cousin with a
different dispersion
relation~\onlinecite{guwen2,PhysRevD.81.104033}. So far these are
the only three types of known stable ABL states with emergent
gapless bosonic excitations without assuming any symmetry. The
Bose metal phase proposed in
Ref.~\onlinecite{balentsfisher,xufisher} rely on a special quasi
one dimensional conservation, which is different from the
scenarios we will focus on.

In this work, we expand these ideas even further, demonstrating
that there are an infinite number of gapless phases that fit into
this class of states. We provide several examples of these
so-called ``higher-rank" ABL theories. We also investigate the
topological properties of these models, showing that they are
``topologically ordered" in the same sense as the photon and
graviton theories, even though they are gapless in the bulk. At
finite system size $L$, the emergent gauge bosons will lead to an
energy splitting between different sectors that scales as a power
law of $1/L$.


\section{Review of Rank-1 and -2 Theories}

\subsection{The Rank-1 Case}

We first review the essential facts about the well-known $U(1)$
photon ABL phase in $3+1d$. In order to connect to the more
general construction, we will address the problem from a somewhat
different (but physically equivalent) viewpoint than the original
works~\cite{PhysRevB.69.064404,wen2003,sondhiphoton}. The gauge
structure (and its duality) is of paramount importance, so we will
omit some details in favor of a more easily generalizable
procedure. For simplicity, we will consider the cubic lattice,
where spins are defined on the links, i.e. the corner-sharing
octahedra. The most important term of the Hamiltonian is simply an
Ising antiferromagnetic interaction on each octahedron: \beqn H =
\frac{J}{2}\sum_{oct} (S^z_{oct})^2, \label{J}\eeqn With a very
large $J$, this term will give rise to a locally conserved
$z$-component of spin, which we will enforce as a constraint on
the low energy Hilbert space: \beqn \sum_{i \in oct}S^z_i = 0.
\label{constraint}\eeqn There are certainly other terms on the
lattice that involve $S^{\pm}$, but their specific forms are not
important, as long as they are all dominated by the $J$ term.
Under the standard change of variables $S^z \sim n - 1/2$ and
$S^\pm \sim e^{\pm i\theta}$, this model becomes a boson rotor
model on the links of a cubic lattice. Noting that the locally
conserved integer $S_{oct}^z$ generates a $U(1)$ gauge symmetry,
after we change variables again to $E_{rr'}\sim (-1)^r n_{rr'}$
and $A_{rr'} \sim (-1)^r \theta_{rr'}$. Note that $A_{rr'}$ is
only defined modulo $2\pi$.

The operators $E$ and $A$ are defined on the links of the lattice,
and this endows them with a vector structure.  We can thus
identify a vector of operators $\textbf{E}(\vec{x})$ and
$\textbf{A}(\vec{x})$ at each site of the cubic lattice, along
with lattice derivatives $\partial_i E_j(\vec{x}) =
E_j(\vec{x}+\hat{i}) - E_j(\vec{x})$. They satisfy the normal
commutation relations $[A_j(\vec{x}), \ E_k(\vec{y})] =
i\delta_{jk}\delta^3(\vec{x} - \vec{y})$.  When phrased in terms
of these new variables, the low energy effective Hamiltonian is
bound to take the following form: \beqn H =
U\sum_{r}\textbf{E}(r)^2 -
K\sum_{\Box}\cos[\text{curl}(\textbf{A})_{\Box}] \eeqn Once we
project all the physics down to the low energy subspace of the
Hilbert space that obeys the constraint imposed by the $J$ term,
the low energy effective Hamiltonian must have the gauge symmetry
$\textbf{A} \rightarrow \textbf{A} + \mathbf{\nabla} f$, which is
generated by the local constraint which can now be written as
\beqn \label{lc1}
\partial_i E_i = 0. \eeqn

Tentatively ignoring the fact that $\textbf{A}$ is compactly
defined, we can expand the low energy effective Hamiltonian at the
minimum of the cosine function (spin wave expansion): \beqn
\label{h1} H = U\sum_r E_i^2 + \frac{K}{2} \sum_r
(\epsilon_{ijk}\partial_j A_k)^2 \eeqn Equation (\ref{h1}) is the
effective low energy Hamiltonian for $(3+1)d$ quantum
electrodynamics (QED) in its deconfined phase. Solving the
equation of motion of Eq.~\ref{h1} from the Heisenberg equation
directly, we will obtain a gapless photon excitation with linear
dispersion relation $\omega \sim c |\vec{k}|$, where the speed of
light $c \sim \sqrt{UK}$. However, we know that in $2+1d$, the
compact QED suffers from the instanton effect: proliferation of
magnetic monopoles in the space time opens up the photon gap, but
that effect is only made clear in terms of the dual variables.
Thus, we will also consider the dual theory to ascertain if there
is a similar gap-opening effect.

We see that the solution of $E_i$ to the local constraint
Eq.~\ref{lc1} can be written as the curl of another vector field
$h_i$, $E_i = \epsilon_{ijk}\partial_j h_k$. This new field $h_i$
is defined on each plaquette center and it is canonically
conjugate to the magnetic field $B_i$. We can now rewrite the
Hamiltonian Eq.~\ref{h1} as \beqn \label{dh1} H = U\sum_r
(\epsilon_{ijk}\partial_j h_k)^2 + \frac{K}{2} \sum_r B_i^2 \eeqn

In contrast to the $2+1d$ case, this new Hamiltonian has the same
form as the original Eq. (\ref{h1}), and formally $h_i$ has the
same gauge symmetry as $A_i$: \beqn h_i \rightarrow h_i + \nabla_i
f. \label{dualgauge} \eeqn We thus say that the system (at least
in the photon phase) is \textit{self-dual}. This is an emergent
feature in the infrared.

In the dual theory, we might expect relevant ``vertex operators"
$\alpha \cos(2\pi N h_i)$, whose analogue in $(2+1)d$ plays the
role of the flux creation. In $(3+1)d$, this vertex operator
corresponds to hopping of the magnetic monopole of the compact
U(1) gauge field. Whether this vertex operator is important or
not, can be determined by evaluating its correlation function in
the limit where $\alpha = 0$. However, in $(3+1)d$, the
correlation function between two such terms in the limit $\alpha =
0$ is \beqn \langle \cos(2\pi N h_i(\vec{x})) \ \cos(2\pi N
h_j(\vec{y}))\rangle_0 \sim \delta_{ij}\delta^3(\vec{x} -
\vec{y}), \eeqn because this is not a gauge invariant correlation
function under gauge transformation Eq.~\ref{dualgauge}. Thus at
the Gaussian fixed point, the vertex operators cos$(2\pi N h_i)$
are irrelevant - at least perturbatively. When the vertex operator
is strong enough, it will induce magnetic monopole condensation
and drive the system into the confined phase. Thus, the gapless
photon is perturbatively protected by the gauge symmetry of both
the original and the dual theory, i.e. the self-duality protects
the stability of the photon phase.

Our review in this subsection is no more than restating the known
fact that the $(3+1)d$ compact U(1) gauge field has a deconfined
phase, which corresponds to the phase where neither the charge nor
the magnetic monopole condenses. In this subsection, we identified
the photon phase where the magnetic monopole is gapped as the
phase where the vertex operator is irrelevant. The language and
logic used in this subsection can be conveniently generalized to
other ABL phases.

\subsection{The Rank-2 Case}

In this section we will review the construction of another ABL
phase with rank-2 tensor gapless bosonic excitations that are
analogous to gravitons. We will omit the exact details of the
microscopic derivation; interested readers are referred to the
original papers
Ref.~\onlinecite{2006cond.mat..2443X,PhysRevB.74.224433}.

The $3+1d$ microscopics in this system give rise to a symmetric
rank-2 tensor field, in contrast to the rank-1 tensor field in the
previous example.  Once again, the system can be simply phrased in
terms of the boson number $n_{ij}$ and its canonical conjugate
phase variables $\theta_{ij}$. We can again define gauge field
variables $E_{ij} \sim n_{ij}$ ($i \neq j$), $E_{ii} \sim
2n_{ii}$, and $A_{ij} \sim \theta_{ij}$, noting that $A_{ij}$ is
compactly defined with modulo $2\pi$.

The low-energy subspace has the local constraint\beqn \label{lc2}
\partial_i E_{ij} = 0, \eeqn which is imposed by a large local
term similar to Eq.~\ref{J}. This constraint generates the gauge
transformation \beqn A_{ij} \rightarrow A_{ij} +
\frac{1}{2}\left(\partial_i \lambda_j + \partial_j
\lambda_i\right) \eeqn This gauge transformation is the same as
that of linearized gravity, if we were to treat $A_{ij}$ as the
fluctuation of a background metric $\eta_{ij}$.  Hence, we term
the gauge boson a ``graviton". The original
works~\cite{2006cond.mat..2443X,PhysRevB.74.224433} use the
language of general relativity to write the Hamiltonian in terms
of the curvature tensor for $A_{ij}$.  We will avoid that notation
here while noting that it has very nice connections to the
Lifshitz gravity proposed independently in
Ref.~\onlinecite{horava2008,horava2009,horava2009a}.

We want to establish the simplest Hamiltonian possible that is
gauge-invariant.  To do this, we will again write down a
gauge-invariant quantity $B_{ij}$ which should be thought of as
the ``2-curl" of $A_{ij}$: \beqn \label{b2} B_{ij} =
\epsilon_{iab}\epsilon_{jcd}\partial_a\partial_c A_{bd}\eeqn

The low energy effective Hamiltonian, or the Hamiltonian after the
``spin-wave expansion", then takes the simple form \beqn H =
U\sum_r E_{ij}^2 + K\sum_r B_{ij}^2,\eeqn where $U$ and $K$ may,
in general, take different values for the $\sum_{ij}X_{ij}^2$ and
$\sum_i X_{ii}^2$ terms, if an ordinary cubic lattice symmetry is
assumed; however, lattice symmetry is not essential to our work
here.

The spin-wave expanded Hamiltonian above already gave us a gapless
``graviton-like" bosonic mode with a quadratic dispersion. In
order to guarantee that this gapless mode is not ruined by the
compactness of the gauge field, we must once again consider the
dual theory. The dual variables solve the constraint equation
(\ref{lc2}), and we can write $E$ as the 2-curl of a new field
$h$: \beqn E_{ij} =
\epsilon_{iab}\epsilon_{jcd}\partial_a\partial_c h_{bd}. \eeqn We
see that $h$ transforms under the same gauge transformation as
tensor $A$ and is canonically conjugate to the tensor field
$B$~\cite{2006cond.mat..2443X,PhysRevB.74.224433}. The vertex
operators will take the form $\cos(2\pi N h_{ij})$, and just as
before the gauge-dependence makes them irrelevant at the infrared
Gaussian fixed point because it violates the gauge symmetry of the
Gaussian fixed point field theory. This will once again guarantee
the gaplessness of the graviton mode, and we see from the above
Hamiltonian that $\omega \sim k^2$.

\subsection{Additional constraints}

For $n \geq 2$, the rank-$n$ theories have additional structure
because they can accommodate several types of local constraints.
Interestingly, we can also enforce more than one local constraint
simultaneously.  For example, we can take the above theory and
additionally require that \beqn \label{lc3} E = \sum_i E_{ii} = 0
\eeqn This generates the gauge transformation \beqn A_{ij}
\rightarrow A_{ij} + \delta_{ij}\lambda \eeqn We now ask a
modified question - what is the simplest theory that is invariant
under the gauge transformations generated by constraints
(\ref{lc2}) and (\ref{lc3}) simultaneously?  We see that our
definition of $B_{ij}$ in Eq. (\ref{b2}) is not good enough.
However, we can use the quantity $B = \sum_i B_{ii}$ to define a
new tensor: \beqn \label{q2} Q_{ij} = \epsilon_{ikl}\left(B_{jl} -
\frac{1}{2}\delta_{jl}B\right), \eeqn which is invariant under
both gauge transformations.  The new effective low energy
Hamiltonian is now~\cite{PhysRevD.81.104033}: \beqn H = U\sum_r
E_{ij}^2 + K\sum_r Q_{ij}^2. \label{z3}\eeqn The new dual fields
are defined in the same way, where $E$ and $h$ have the same
functional relation as $Q$ and $A$~\cite{PhysRevD.81.104033}. Thus
this theory is again ``self-dual" with identical gauge symmetries
on the two sides of the duality. This theory (Eq.~\ref{z3}) is
again gapless, though it has a different dispersion: because there
are now three spatial derivatives of $A$ in the $Q$ tensor
Eq.~\ref{q2}, the dispersion of the low energy excitation is
$\omega \sim k^3$.

\section{General Procedure}



To generalize these arguments to higher rank tensor fields, we
need to first establish which types of gauge transformations will
be allowed. To simplify our discussion, we want the field theory
to be rotationally symmetric, though it is possible that the
lattice regularization may possess irrelevant rotation-breaking
terms. Additionally, the gauge constraint should depend only on
$E_{ijk...}$ and no other locally defined tensor fields.

These two requirements restrict the constraints that we will
consider to higher-dimensional versions of the Gauss law and
traceless conditions.  These constraints are ``rotationally"
symmetric in the correct way to respect lattice symmetries (again,
we stress that the states we construct should be insensitive to
weak lattice symmetry breaking). We enumerate the allowed gauge
transformations in Table 1 for rank one through three.  To
simplify notation, we denote the symmetrizing operation

\beqn
T_{(ijk)} = \frac{1}{3!}\left(T_{ijk} + T_{jik} + sym\right)
\eeqn

\begin{center}
\begin{table}
\begin{tabular}{| l | l | l |}
\hline
Rank of theory & Local constraint & Gauge transformation \\ \hline
\multirow{1}{*}{$n=1$} & $\partial_i E_i = 0$ & $A_i \rightarrow A_i + \partial_i \lambda$ \\ \hline

\multirow{3}{*}{$n=2$} & $\partial_i E_{ij} = 0$ & $A_{ij} \rightarrow A_{ij} + \partial_{(i}\lambda_{j)}$ \\
& $\partial_i \partial_j E_{ij} = 0$ & $A_{ij} \rightarrow A_{ij} + \partial_i \partial_j \lambda$ \\
& $E_{ii} = 0$ & $A_{ij} \rightarrow A_{ij} + \delta_{ij}\lambda$ \\ \hline

\multirow{5}{*}{$n=3$} & $\partial_i E_{ijk} = 0$ & $A_{ijk} \rightarrow A_{ijk} + \partial_{(i}\lambda_{jk)}$* \\
& $\partial_i \partial_j E_{ijk} = 0$ & $A_{ijk} \rightarrow A_{ijk} + \partial_{(i}\partial_j\lambda_{k)}$ \\
& $\partial_i \partial_j \partial_k E_{ijk} = 0$ & $A_{ijk} \rightarrow A_{ijk} + \partial_i\partial_j\partial_k\lambda$ \\
& $\delta_{ij} E_{ijk} = 0$ & $A_{ijk} \rightarrow A_{ijk} + \delta_{(ij}\lambda_{k)}$* \\
& $\delta_{ij}\partial_k E_{ijk} = 0$ & $A_{ijk} \rightarrow A_{ijk} + \delta_{(ij}\partial_{k)}\lambda$ \\ \hline

\end{tabular}
\caption{Allowed gauge transformations which are rotationally invariant and do not depend on an auxiliary tensor field.\\
* These gauge transformations are not totally independent - $\lambda_{jk}$ should be made traceless.}
\end{table}
\end{center}

An important generic question is the number of gapless modes in
the system. This is determined by switching to a Lagrangian
formulation and thinking of the $\lambda$ tensor as a Lagrange
multiplier. Each degree of freedom of $\lambda$ will reduce the
number of gapless modes by one (though there is a subtlety to this
counting, which is detailed in the appendix). For example, for the
familiar photon phase, \beqn L_1 = E^2 - B^2 + \lambda(\partial_i
E_i). \eeqn $E_i$ has three components initially, so the one free
component of a scalar $\lambda$ reduces the number of gapless
modes to the familiar two of the photon. For higher rank cases,
though it quickly becomes tedious to count the number of free
components of an arbitrary rank symmetric tensor, the idea is
straightforward. Indeed, it is also possible to diagonalize the
Hamiltonian directly, and this reproduces the previous results.

The essential component of many ABL theories is the process by
which gap-opening perturbations are prohibited.  Generically, any
relevant term in the Lagrangian should open a gap, and so to
eliminate \textit{all} such terms places strict requirements on
the theory. In the theories we consider in this paper, we use
gauge-invariance and self-duality to protect the photon gap from
perturbations at a Gaussian IR fixed point, just like the examples
reviewed in the previous section.

The gauge structure in all of the theories we consider is emergent
in the IR. Indeed, it is due to a constraint on the low-energy
Hilbert space of the microscopic model. This means that the
gapless phase is not stable to arbitrarily strong perturbations,
since moving out of the constrained subspace generically destroys
the gauge structure. As an example, consider the gauge charge
excitation in the rank-1 theory. The low-energy subspace is that
of the charge vacuum, but if we tune the charge gap to zero, the
gauge charges condense and gap out the gauge boson through the
Higgs mechanism.

Additionally, the gauge structure will constrain the form of the
Hamiltonian. As we have seen above, we want to use $A$ to
construct two gauge-invariant tensors $E$ and $B$ which play the
usual roles in electromagnetism. Given that there is a direct
relation between the gauge transformations on $A$ and the
constraints on $E$, it is a straightforward task to build the most
relevant terms. In this case, ``most relevant'' means that $B$ has
the fewest number of spatial derivatives of $A$, but it must be
gauge invariant still.

Just to limit the variety of states, we require rotational
invariance in this paper, which also constrains the form of the
Hamiltonian (as does gauge invariance). But we want to stress that
weakly breaking the rotational invariance will not destroy the
states we construct, namely it will not gap out the bosonic modes
of the ABL phase. For example, the low energy photon excitations
of the ABL phase studied in
~\onlinecite{PhysRevB.69.064404,wen2003,sondhiphoton} have a
rotational invariant dispersion at low energy, but we know that
breaking the rotational invariance will not destroy the photon
excitations. The local gauge constraints are similarly influenced
by the requirement of rotational invariance, as was noted above.
We can then consider tensor representations of rotational group
$SO(3)$, and it turns out that we will only be interested in the
symmetric pieces.

For example, to construct the gauge invariant rank-3 magnetic
field $Q_{ijk}$ with both a derivative constraint and a trace
constraint on $E_{ijk}$, the resulting theories will involve
$B_{ijk}$ which is a 3-curl of $A$ and $B_k = B_{iik}$. Because
$Q_{ijk}$ carries three vector indices, it can be constructed with
three vector representation of $SO(3)$. The standard expansion of
a tensor defined over three copies of the fundamental
representation of $SO(3)$ is

$$1 \otimes 1 \otimes 1 = 3 \oplus 2 \oplus 2 \oplus 1 \oplus 1 \oplus 1 \oplus 0$$

The spin-2 and spin-0 pieces here are antisymmetric in at least
two indices. Requiring overall symmetrization will reduce the
expansion to a fully symmetric spin-3 part $T_{(ijk)}$ and a
symmetric spin-1 part $T^{\prime}_{ijk} = \delta_{(ij}T_{k)}$.
Thus, we can understand connection between the allowed constraints
and how they will involve traces of the curls of $A$ by
considering which parts of the tensor representation are
symmetric.

However, as was noted previously, gauge structure is not enough to
guarantee the gaplessness of the photon.  In 2+1d this manifests
as the so-called ``instanton effect'', which is to say that the
magnetic flux insertion operator is always relevant at the
Gaussian fixed point. Thus, the instantons proliferate and open a
gap for the photons. Thus in general in our $(3+1)d$ ABLs, we need
to argue that all of the vertex operators that generically take
the form $\cos\left(2\pi N h_{\alpha\beta...}\right)$ for the dual
gauge field $h$ are irrelevant.

\section{Examples}

In this section we will discuss a few examples of new ABL phases.
The first example is similar to the graviton theories detailed in
the previous section, except that it has a different local
constraint. This is an interesting property of rank-$n$ theories
for $n \geq 2$ which greatly enhances the variety of gapless gauge
theories. There are roughly $n$ different constraints involving
only derivatives for a given rank-$n$ theory in addition to the
various types of traceless conditions.

The original graviton model had as its local constraint
Eq.~\ref{lc2}.  We can instead contract another derivative on
$E_{ij}$ to get a different theory: \beqn \label{lc4}
\partial_i \partial_j E_{ij} = 0
\eeqn Compared to the theory governed by Eq.~\ref{lc2}, this
theory has a scalar (as opposed to vector) charge and has five
total degrees of freedom (up from three).  Even before determining
the simplest possible Hamiltonian, we see that the gapless
excitations are distinct in character from the original gravitons:
\beqn A_{ij} \rightarrow A_{ij} + \partial_i\partial_j \lambda
\eeqn


We can construct the Hamiltonian of this ABL state using the
following symmetrized gauge invariant tensor field $B$: \beqn
\label{b3} B_{ij} = \frac{1}{2}\left(\epsilon_{iab}\partial_a
A_{bj} + \epsilon_{jcd}\partial_c A_{id}\right). \eeqn The
corresponding low energy Hamiltonian again takes the schematic
form of $E^2 + B^2$, as before, and this theory is again
self-dual, but it now has a linear dispersion $\omega \sim k$.

We can also consider enforcing the constraint Eq. (\ref{lc3}) in
addition to Eq. (\ref{lc4}).  However, in this case, $B_{ij}$
given by Eq. (\ref{b3}) is already invariant under both gauge
transformations.  In fact, in conjunction with the graviton theory
discussed previously, we have now characterized \textit{all}
rank-2 symmetric gauge theories whose gauge transformations
satisfy our criteria above.


While the rank-1 and rank-2 systems have nice interpretations as
``photons" and ``gravitons" due to the familiarity with known
systems, there is no such nice identification for the rank-3 case.
We cannot leverage any analogy to linearized gravity nor
electromagnetism, and instead we will proceed using our general
method.


To illustrate this case, we consider two canonically conjugate
symmetric rank-3 tensor fields $A_{ijk}$ and $E_{ijk}$ where $A$
is defined modulo $2\pi$.  We then impose the local constraint
\beqn \label{lc5}
\partial_i E_{ijk} = 0
\eeqn which generates the gauge transformation \beqn \label{gt3} A_{ijk}
\rightarrow A_{ijk} + \partial_{(i}\lambda_{jk)} \eeqn The
corresponding lattice system is given in the appendix.  As before,
we seek a ``magnetic'' field that is gauge-invariant and of lowest
number of derivatives of $A$.  Additionally, it should be
symmetric.  We see that \beqn B_{ijk} =
\epsilon_{iab}\epsilon_{jcd}\epsilon_{kef}\partial_a\partial_c\partial_e
A_{bdf} \label{B-rank3}\eeqn is the simplest tensor that fits the
requirements. From this tensor we can construct a state with the
following low energy effective Hamiltonian \beqn H = U\sum_r E^2 +
K\sum_r B^2, \eeqn where the coefficients of the $\sum X_{iii}^2$,
$\sum X_{iij}^2$, and $\sum X_{ijk}^2$ terms may in general be
different.  This system is self-dual in the same way as before, by
defining the dual variable $h_{ijk}$ as \beqn E_{ijk} =
\epsilon_{iab}\epsilon_{jcd}\epsilon_{kef}\partial_a\partial_c\partial_e
h_{bdf} \eeqn and requiring that $h_{ijk}$ transform in the same
way as $A_{ijk}$ under a change of gauge.  The vertex operators of
the dual variables $\cos\left(2\pi N h_{ijk}\right)$ are easily
seen to be gauge dependent, and thus irrelevant in the same way as
before.  This is a new gapless Bose liquid with $\omega \sim k^3$
with four independent modes for each momentum $\vec{k}$.

We can then ask what happens when another local constraint is
imposed.
\begin{equation}\label{lc6}
\delta_{ij} E_{ijk} = 0.
\end{equation}
This constraint gives rise to the gauge transformation \beqn
A_{ijk} \rightarrow A_{ijk} + \delta_{(ij}\lambda_{k)}, \eeqn
which provides a nice example of the ``mode overcounting''
discussed in the appendix.  In particular, the above constraint
gives rise to new physics only when the 1-form field $\lambda_k$
is not exact, $i.e.$ $\lambda_k \neq
\partial_k \Gamma$. If $\lambda_k$ is a total derivative of some
scalar function, then this constraint Eq.~\ref{lc6} is not
independent of the transformation Eq.~\ref{gt3} and the system as described by $B_{ijk}$ given before
in Eq.~\ref{B-rank3} is invariant under both.

If $\lambda_k \neq \partial_k \Gamma$, then we have to construct a
new ``magnetic field" that is invariant under both gauge
transformations.  To do so, we need to define two quantities:
\beqn
D_{ij} = \delta_{ij}\partial^2 - \partial_i\partial_j \\
B_k = B_{iik}. \eeqn Using these quantities, the new low energy
effective Hamiltonian is schematically $E^2 + Q^2$ where we have
defined \beqn Q_{ijk} = \partial^2 B_{ijk} -
\frac{3}{4}D_{(ij}B_{k)}. \eeqn This theory has a rather soft
dispersion $\omega \sim k^5$, and only one single mode at each
momentum $\vec{k}$. And just like all the examples before, this
theory is also self-dual.

Continuing this procedure to higher rank theories generates an
entire infinite family of ABL phases.  The procedure is exactly
the same, though the precise enumeration of possible gauge
transformations (and, indeed, even the number of degrees of
freedom) becomes tedious quickly.  However, by leveraging the
gauge structure in addition to the self-duality at the IR fixed
point, we are able to in all cases derive the appropriate
low-energy effective Hamiltonian for a given local constraint.

\section{Topological Order}

The $U(1)$ spin liquid in $3+1d$, in addition to its stability,
also possesses a curious type of topological order, which was
discussed in detail in Ref.~\onlinecite{PhysRevB.69.064404}. When
the system is put on a three dimensional torus $T^3$ with size
$L^3$, it is possible to thread electric flux around each of the
noncontractible loops. The flux integrals each commute with the
low-energy Hamiltonian and each other, so they constitute
constants of motion. When the flux spreads out over the whole
thermodynamically large system, the energy cost goes to zero as
$1/L$.  An identical picture holds for the magnetic flux, so
topological order is characterized by six integers. The system is
stable due to the gap in both electric and magnetic charges,
which makes it exponentially unlikely for a ``particle-hole''
pair to be created and propagate all the way around the torus to
change the flux.

There is a similar construction for the graviton ABL discussed in
Ref.~\onlinecite{2006cond.mat..2443X,PhysRevB.74.224433}. However,
one must be more careful in the selection of which fluxes of
$E_{ij}$ are used.  In principle, there are twenty seven different
fluxes - three orientations of the flux surface and nine
components of $E_{ij}$.  Upon calculation, one can show that
fifteen of these are zero, and of the remaining twelve only nine
are independent.  The same result holds for the magnetic fluxes,
meaning that the graviton ABL has topological order characterized
by eighteen integers.  It is similar to the $U(1)$ case in that
the ground states are split by $1/L$ and are exponentially
unlikely to mix.

We claim that similar arguments hold for the whole infinite family
of theories constructed in the previous sections with only
derivative constraints.  The topological order is characterized by
$6k$ integers corresponding to the electric and magnetic fluxes,
where $k$ the number of independent components of the charge
tensor.  Since the Hamiltonian densities for these theories are
generically $E^2 + B^2$, we expect that in all cases the ground
state degeneracy closes as $1/L$ in the thermodynamic limit.


To understand the origin of the $6k$, we consider a generic local
constraint written as \beqn (\partial_i \hat{E}_{ijk...} -
\hat{\rho}_{jk...})|Phys\rangle = 0 \eeqn It is natural to
interpret the violations of the local constraints as ``charges.''
The particular choice of contraint endows the charges with a
tensor structure, and the underlying symmetry of $E_{ijk...}$ is
reflected in that structure.  Going back to the lattice model,
these charges can also be thought of as the open ends of strings.
The constraint is then interpreted as the condition that strings
do not end on sites.  Due to the all-important electromagnetic
duality protecting these phases, there is a corresponding magnetic
charge tensor with exactly the same structure as the electric
charge.

Using these charge tensors, we can then create ``particle-hole"
pairs of a given type of charge and wind them around a
noncontractible loop of $T^3$. Three dimensions times two species
of charge gives the factor of six, and there are $k$ independent
charges depending on the particular constraint. We see that $k=1$
for the ordinary QED, while $k=3$ for the graviton ABL which has a
vector charge.

To extend these ideas to constraints with more derivatives, we see
that the constraint $\partial_i \partial_j E_{ij} = 0$ can be
rewritten as $\partial_i F_i = 0$ for a vector field $F_i =
\partial_j E_{ij}$.  This constraint has a scalar charge, and it
can be shown easily that the fluxes of $F_i$ commute with each
other and the Hamiltonian.  This extra step is needed to invoke
the divergence theorem, since we need to work with the divergence
of a vector field.  Importantly, the characterization is still the
same - since the underlying charge is a scalar, this theory is
characterized by six winding numbers.

Finally, we consider the second type of constraint detailed above,
such as $\delta_{ij} E_{ij} = 0$. In the rank-2 case, we imagine
threading a flux of $E_{xx}$ around the noncontractible loop in
the $x$-direction while simultaneously threading a flux of
$E_{yy}$ in the $y$-direction.  The two ``strings'' involved in
this threading process need not intersect, but in the ground state
the fluxes spread out over the whole system.  Once this occurs, we
see that the traceless constraint fixes the flux of $E_{zz}$
through the $z$-direction so that the three integers sum to zero.
This new phase is characterized by 16 integers.  Extending these
constraints to higher-rank theories is straightforward but
tedious, and simply removes topological degrees of freedom from
the diagonal fluxes.

\section{Summary and Discussion}



In this work we have demonstrated that there is an infinite family
of strongly-correlated gapless boson systems whose low-energy
Hilbert space does not break any symmetries with gapless
excitations stable with respect to small arbitrary perturbations.
The gaplessness is protected by a combination of emergent gauge
invariance (enforced by a local constraint on the low-energy
Hilbert space) and a generalized electromagnetic duality.  Within
some limitations, the dispersion and representation of the
emergent gauge boson can be tuned. Additionally, these theories
have an interesting type of topological order characterized by
$6k$ integers, depending on the exact underlying local constraint.

Although we have made heavy use of the gauge structure in
constructing these ABL phases, we have not made a careful analysis
of the associated gauge groups.  Apart from the simplest $U(1)$
spin liquid, the higher-rank gauge fields are not algebra-valued
1-forms, and therefore do not fit into the standard Yang-Mills
architecture.

Our rank-2 model with constraint $\partial_i E_{ij} = 0$ can
potentially be thought of as $U(1) \times U(1) \times U(1)$ gauge
theory after an (unusual) symmetrization between the space index and
the flavor index; this provides a possible realization of our
rank-2 theory starting with three copies of the photon phase. We also
note that this connection between linearized gravity and the Yang-Mills
gauge theory was already observed in the loop quantum gravity
literature~\cite{PhysRevD.66.024017,Contreras:2013qua}.  Further study of
these models will hopefully elucidate these connections.

\section*{Acknowledgments}

The authors are supported by the David and Lucile Packard Foundation and NSF Grant No. DMR-1151208.
The authors thank Xiao-Gang Wen for helpful online discussions.  AR
would like to thank Zhen Bi and Dominic Else for helping sharpen several arguments.

\section*{Appendix A - Rank-3 Lattice Hamiltonian}

For concreteness, we will present a lattice Hamiltonian for one of
the rank-3 cases.  In line with the previous work by Xu, this
Hamiltonian has two pieces: a generic boson hopping term and a
density-density repulsion term.


The unit cell for this lattice consists of a face-centered cubic
lattice that also has a site at the center (see Figure 1).  The boson occupation
at the corners of the fcc unit cell are three-fold degenerate,
labeled $n_{xxx,\vec{r}}$, $n_{yyy,\vec{r}}$, and
$n_{zzz,\vec{r}}$.  The faces are two-fold degenerate with labels
$n_{xxy,\vec{r}+\hat{x}/2+\hat{y}/2}$,
$n_{xyy,\vec{r}+\hat{x}/2+\hat{y}/2}$, and so on, and the center
is labeled $n_{xyz,\vec{r}+\hat{x}/2+\hat{y}/2+\hat{z}/2}$.

\begin{figure}[!htb]
\centering
\includegraphics[scale=0.8]{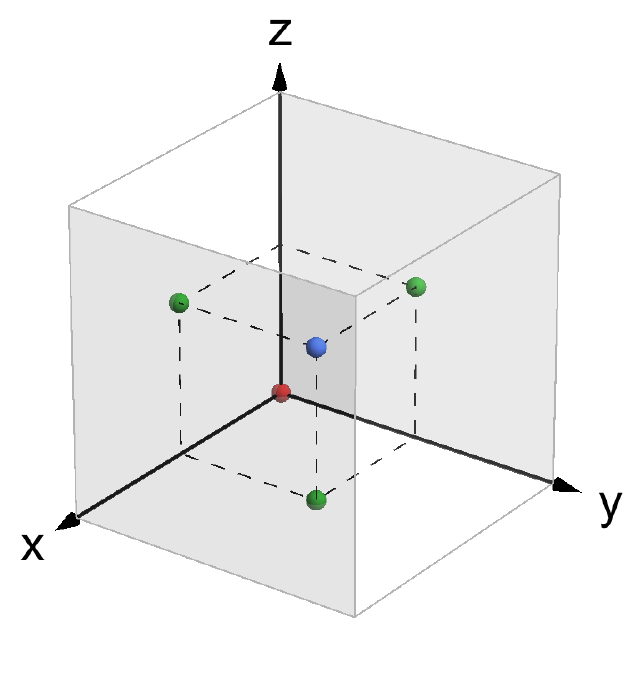}
\caption{The unit cell for the simplest rank-3 model.  The red site is three-fold degenerate ($n_{xxx}$), the green sites are two-fold degenerate ($n_{xxy}$) and the blue site is nondegenerate ($n_{xyz}$).}
\end{figure}

The hopping term of the Hamiltonian $H = H_t + H_v$ is generic,
and in principle contains all 45 exchanges.  The potential takes
the form for average boson density $\bar{n}$

\beqn
H_v = H_{xx} + H_{yy} + H_{zz} + H_{xy} + H_{yz} + H_{xz}
\eeqn

\begin{multline}
H_{xx} = V(n_{xxx,\vec{r}}+n_{xxx,\vec{r}+\hat{x}}+n_{xyx,,\vec{r}+\hat{x}/2+\hat{y}/2}+\\
n_{xyx,\vec{r}+\hat{x}/2-\hat{y}/2}+n_{xzx,\vec{r}+\hat{x}/2+\hat{z}/2}+\\
+n_{xzx,,\vec{r}+\hat{x}/2-\hat{z}/2} - 6\bar{n})^2
\end{multline}

\begin{multline}
H_{xy} = V(n_{xyx,\vec{r}+\hat{x}/2+\hat{y}/2}+n_{xyx,\vec{r}+3\hat{x}/2+\hat{y}/2}+ \\
+n_{xyy,\vec{r}+\hat{x}/2+\hat{y}/2}+n_{xyy,\vec{r}+\hat{x}/2+3\hat{y}/2}+ \\
+n_{xyz,\vec{r}+\hat{x}/2+\hat{y}/2+\hat{z}/2}+n_{xyz,\vec{r}+\hat{x}/2+\hat{y}/2+3\hat{z}/2} - 6\bar{n})^2
\end{multline}

with similar expressions for the other four terms.  We define
$E_{ijk} = (-1)^{\vec{r}}(n_{ijk} - \bar{n})$ and use the usual
lattice derivative to see that the low-energy subspace of this
Hamiltonian has the local constraint $\partial_i E_{ijk} = 0$.

\section*{Appendix B - Mode Overcounting}

There is a subtle point that needs to be addressed about the
definitions of the above gauge transformations for rank greater
than or equal to three.  In particular, the two gauge
transformations below are not totally independent:

\beqn
A_{ijk} \rightarrow A_{ijk} + \partial_{(i}\lambda_{jk)}
\\
A_{ijk} \rightarrow A_{ijk} + \delta_{(ij}\partial_{k)}\lambda
\eeqn

The first contains the second as a special case.  This is
understood in terms of tensor representations of $SO(3)$ by noting
that a symmetric rank-2 tensor (six degrees of freedom) has a
single scalar trace mode in addition to the five spin-2 modes.  As
such, a more correct accounting would require tracelessness of
$\lambda_{ij}$, which is achieved via

\beqn
\tilde{\lambda}_{ij} = \lambda_{ij} - \frac{1}{3}\delta_{ij}\lambda_{kk}
\eeqn

This type of overcounting of trace modes persists into higher
rank, and becomes increasingly complicated as the number of trace
modes increases.

\bibliography{ref}

\begin{thebibliography}{25}
\expandafter\ifx\csname natexlab\endcsname\relax\def\natexlab#1{#1}\fi
\expandafter\ifx\csname bibnamefont\endcsname\relax
  \def\bibnamefont#1{#1}\fi
\expandafter\ifx\csname bibfnamefont\endcsname\relax
  \def\bibfnamefont#1{#1}\fi
\expandafter\ifx\csname citenamefont\endcsname\relax
  \def\citenamefont#1{#1}\fi
\expandafter\ifx\csname url\endcsname\relax
  \def\url#1{\texttt{#1}}\fi
\expandafter\ifx\csname urlprefix\endcsname\relax\def\urlprefix{URL }\fi
\providecommand{\bibinfo}[2]{#2}
\providecommand{\eprint}[2][]{\url{#2}}

\bibitem[{\citenamefont{Hermele et~al.}(2004)\citenamefont{Hermele, Fisher, and
  Balents}}]{PhysRevB.69.064404}
\bibinfo{author}{\bibfnamefont{M.}~\bibnamefont{Hermele}},
  \bibinfo{author}{\bibfnamefont{M.~P.~A.} \bibnamefont{Fisher}},
  \bibnamefont{and} \bibinfo{author}{\bibfnamefont{L.}~\bibnamefont{Balents}},
  \bibinfo{journal}{Phys. Rev. B} \textbf{\bibinfo{volume}{69}},
  \bibinfo{pages}{064404} (\bibinfo{year}{2004}),
  \urlprefix\url{http://link.aps.org/doi/10.1103/PhysRevB.69.064404}.

\bibitem[{\citenamefont{Wen}(2003)}]{wen2003}
\bibinfo{author}{\bibfnamefont{X.-G.} \bibnamefont{Wen}},
  \bibinfo{journal}{Phys. Rev. B} \textbf{\bibinfo{volume}{68}},
  \bibinfo{pages}{115413} (\bibinfo{year}{2003}).

\bibitem[{\citenamefont{Castelnovo et~al.}(2008)\citenamefont{Castelnovo,
  Moessner, and Sondhi}}]{spinice1}
\bibinfo{author}{\bibfnamefont{C.}~\bibnamefont{Castelnovo}},
  \bibinfo{author}{\bibfnamefont{R.}~\bibnamefont{Moessner}}, \bibnamefont{and}
  \bibinfo{author}{\bibfnamefont{S.~L.} \bibnamefont{Sondhi}},
  \bibinfo{journal}{Nature} \textbf{\bibinfo{volume}{451}}, \bibinfo{pages}{42}
  (\bibinfo{year}{2008}).

\bibitem[{\citenamefont{Gingras and McClarty}(2014)}]{spinice6}
\bibinfo{author}{\bibfnamefont{M.~J.~P.} \bibnamefont{Gingras}}
  \bibnamefont{and} \bibinfo{author}{\bibfnamefont{P.~A.}
  \bibnamefont{McClarty}}, \bibinfo{journal}{Rep. Prog. Phys.}
  \textbf{\bibinfo{volume}{77}}, \bibinfo{pages}{056501}
  (\bibinfo{year}{2014}).

\bibitem[{\citenamefont{Balents}(2010)}]{slbalents}
\bibinfo{author}{\bibfnamefont{L.}~\bibnamefont{Balents}},
  \bibinfo{journal}{Nature} \textbf{\bibinfo{volume}{464}},
  \bibinfo{pages}{199} (\bibinfo{year}{2010}).

\bibitem[{\citenamefont{Fennell et~al.}(2009)\citenamefont{Fennell, Deen,
  Wildes, Schmalzl, Prabhakaran, Boothroyd, Aldus, McMorrow, and
  Bramwell}}]{spinice2}
\bibinfo{author}{\bibfnamefont{T.}~\bibnamefont{Fennell}},
  \bibinfo{author}{\bibfnamefont{P.~P.} \bibnamefont{Deen}},
  \bibinfo{author}{\bibfnamefont{A.~R.} \bibnamefont{Wildes}},
  \bibinfo{author}{\bibfnamefont{K.}~\bibnamefont{Schmalzl}},
  \bibinfo{author}{\bibfnamefont{D.}~\bibnamefont{Prabhakaran}},
  \bibinfo{author}{\bibfnamefont{A.~T.} \bibnamefont{Boothroyd}},
  \bibinfo{author}{\bibfnamefont{R.~J.} \bibnamefont{Aldus}},
  \bibinfo{author}{\bibfnamefont{D.~F.} \bibnamefont{McMorrow}},
  \bibnamefont{and} \bibinfo{author}{\bibfnamefont{S.~T.}
  \bibnamefont{Bramwell}}, \bibinfo{journal}{Science}
  \textbf{\bibinfo{volume}{326}}, \bibinfo{pages}{415} (\bibinfo{year}{2009}).

\bibitem[{\citenamefont{Jaubert and Holdsworth}(2009)}]{spinice3}
\bibinfo{author}{\bibfnamefont{L.~D.~C.} \bibnamefont{Jaubert}}
  \bibnamefont{and} \bibinfo{author}{\bibfnamefont{P.~C.~W.}
  \bibnamefont{Holdsworth}}, \bibinfo{journal}{Nature Physics}
  \textbf{\bibinfo{volume}{5}}, \bibinfo{pages}{258} (\bibinfo{year}{2009}).

\bibitem[{\citenamefont{Ross et~al.}(2011)\citenamefont{Ross, Savary, Gaulin,
  and Balents}}]{spinice4}
\bibinfo{author}{\bibfnamefont{K.~A.} \bibnamefont{Ross}},
  \bibinfo{author}{\bibfnamefont{L.}~\bibnamefont{Savary}},
  \bibinfo{author}{\bibfnamefont{B.~D.} \bibnamefont{Gaulin}},
  \bibnamefont{and} \bibinfo{author}{\bibfnamefont{L.}~\bibnamefont{Balents}},
  \bibinfo{journal}{Phys. Rev. X} \textbf{\bibinfo{volume}{1}},
  \bibinfo{pages}{021002} (\bibinfo{year}{2011}).

\bibitem[{\citenamefont{Benton et~al.}(2012)\citenamefont{Benton, Sikora, and
  Shannon}}]{spinice5}
\bibinfo{author}{\bibfnamefont{O.}~\bibnamefont{Benton}},
  \bibinfo{author}{\bibfnamefont{O.}~\bibnamefont{Sikora}}, \bibnamefont{and}
  \bibinfo{author}{\bibfnamefont{N.}~\bibnamefont{Shannon}},
  \bibinfo{journal}{Phys. Rev. B} \textbf{\bibinfo{volume}{86}},
  \bibinfo{pages}{075154} (\bibinfo{year}{2012}).

\bibitem[{\citenamefont{Moessner and Sondhi}(2003)}]{sondhiphoton}
\bibinfo{author}{\bibfnamefont{R.}~\bibnamefont{Moessner}} \bibnamefont{and}
  \bibinfo{author}{\bibfnamefont{S.~L.} \bibnamefont{Sondhi}},
  \bibinfo{journal}{Phys. Rev. B} \textbf{\bibinfo{volume}{68}},
  \bibinfo{pages}{184512} (\bibinfo{year}{2003}).

\bibitem[{\citenamefont{{Xu}}(2006)}]{2006cond.mat..2443X}
\bibinfo{author}{\bibfnamefont{C.}~\bibnamefont{{Xu}}},
  \bibinfo{journal}{eprint arXiv:cond-mat/0602443}  (\bibinfo{year}{2006}),
  \eprint{cond-mat/0602443}.

\bibitem[{\citenamefont{Gu and Wen}(2006)}]{guwen1}
\bibinfo{author}{\bibfnamefont{Z.-C.} \bibnamefont{Gu}} \bibnamefont{and}
  \bibinfo{author}{\bibfnamefont{X.-G.} \bibnamefont{Wen}},
  \bibinfo{journal}{arXiv:gr-qc/0606100}  (\bibinfo{year}{2006}).

\bibitem[{\citenamefont{Xu}(2006)}]{PhysRevB.74.224433}
\bibinfo{author}{\bibfnamefont{C.}~\bibnamefont{Xu}}, \bibinfo{journal}{Phys.
  Rev. B} \textbf{\bibinfo{volume}{74}}, \bibinfo{pages}{224433}
  (\bibinfo{year}{2006}),
  \urlprefix\url{http://link.aps.org/doi/10.1103/PhysRevB.74.224433}.

\bibitem[{\citenamefont{Gu and Wen}(2012)}]{guwen2}
\bibinfo{author}{\bibfnamefont{Z.-C.} \bibnamefont{Gu}} \bibnamefont{and}
  \bibinfo{author}{\bibfnamefont{X.-G.} \bibnamefont{Wen}},
  \bibinfo{journal}{Nuclear Physics B} \textbf{\bibinfo{volume}{863}},
  \bibinfo{pages}{90} (\bibinfo{year}{2012}).

\bibitem[{\citenamefont{Xu and Ho\ifmmode~\check{r}\else
  \v{r}\fi{}ava}(2010)}]{PhysRevD.81.104033}
\bibinfo{author}{\bibfnamefont{C.}~\bibnamefont{Xu}} \bibnamefont{and}
  \bibinfo{author}{\bibfnamefont{P.}~\bibnamefont{Ho\ifmmode~\check{r}\else
  \v{r}\fi{}ava}}, \bibinfo{journal}{Phys. Rev. D}
  \textbf{\bibinfo{volume}{81}}, \bibinfo{pages}{104033}
  (\bibinfo{year}{2010}),
  \urlprefix\url{http://link.aps.org/doi/10.1103/PhysRevD.81.104033}.

\bibitem[{\citenamefont{Wen}(1989)}]{wentopo1}
\bibinfo{author}{\bibfnamefont{X.-G.} \bibnamefont{Wen}},
  \bibinfo{journal}{Phys. Rev. B} \textbf{\bibinfo{volume}{7387}},
  \bibinfo{pages}{40} (\bibinfo{year}{1989}).

\bibitem[{\citenamefont{Wen and Niu}(1990)}]{wentopo2}
\bibinfo{author}{\bibfnamefont{X.-G.} \bibnamefont{Wen}} \bibnamefont{and}
  \bibinfo{author}{\bibfnamefont{Q.}~\bibnamefont{Niu}},
  \bibinfo{journal}{Phys. Rev. B} \textbf{\bibinfo{volume}{41}},
  \bibinfo{pages}{9377} (\bibinfo{year}{1990}).

\bibitem[{\citenamefont{Wen}(1990)}]{wentopo3}
\bibinfo{author}{\bibfnamefont{X.-G.} \bibnamefont{Wen}},
  \bibinfo{journal}{Int. J. Mod. Phys. B} \textbf{\bibinfo{volume}{239}},
  \bibinfo{pages}{4} (\bibinfo{year}{1990}).

\bibitem[{\citenamefont{Paramekanti et~al.}(2002)\citenamefont{Paramekanti,
  Balents, and Fisher}}]{balentsfisher}
\bibinfo{author}{\bibfnamefont{A.}~\bibnamefont{Paramekanti}},
  \bibinfo{author}{\bibfnamefont{L.}~\bibnamefont{Balents}}, \bibnamefont{and}
  \bibinfo{author}{\bibfnamefont{M.~P.~A.} \bibnamefont{Fisher}},
  \bibinfo{journal}{Phys. Rev. B} \textbf{\bibinfo{volume}{66}},
  \bibinfo{pages}{054526} (\bibinfo{year}{2002}).

\bibitem[{\citenamefont{Xu and Fisher}(2007)}]{xufisher}
\bibinfo{author}{\bibfnamefont{C.}~\bibnamefont{Xu}} \bibnamefont{and}
  \bibinfo{author}{\bibfnamefont{M.~P.~A.} \bibnamefont{Fisher}},
  \bibinfo{journal}{Phys. Rev. B} \textbf{\bibinfo{volume}{75}},
  \bibinfo{pages}{104428} (\bibinfo{year}{2007}).

\bibitem[{\citenamefont{Ho\v{r}ava}(2009{\natexlab{a}})}]{horava2008}
\bibinfo{author}{\bibfnamefont{P.}~\bibnamefont{Ho\v{r}ava}},
  \bibinfo{journal}{J. High Energy Phys.} \textbf{\bibinfo{volume}{0903}},
  \bibinfo{pages}{020} (\bibinfo{year}{2009}{\natexlab{a}}).

\bibitem[{\citenamefont{Ho\v{r}ava}(2009{\natexlab{b}})}]{horava2009}
\bibinfo{author}{\bibfnamefont{P.}~\bibnamefont{Ho\v{r}ava}},
  \bibinfo{journal}{Phys. Rev. D} \textbf{\bibinfo{volume}{79}},
  \bibinfo{pages}{084008} (\bibinfo{year}{2009}{\natexlab{b}}).

\bibitem[{\citenamefont{Ho\v{r}ava}(2009{\natexlab{c}})}]{horava2009a}
\bibinfo{author}{\bibfnamefont{P.}~\bibnamefont{Ho\v{r}ava}},
  \bibinfo{journal}{Phys. Rev. Lett.} \textbf{\bibinfo{volume}{102}},
  \bibinfo{pages}{161301} (\bibinfo{year}{2009}{\natexlab{c}}).

\bibitem[{\citenamefont{Varadarajan}(2002)}]{PhysRevD.66.024017}
\bibinfo{author}{\bibfnamefont{M.}~\bibnamefont{Varadarajan}},
  \bibinfo{journal}{Phys. Rev. D} \textbf{\bibinfo{volume}{66}},
  \bibinfo{pages}{024017} (\bibinfo{year}{2002}),
  \urlprefix\url{http://link.aps.org/doi/10.1103/PhysRevD.66.024017}.

\bibitem[{\citenamefont{Contreras and Leal}(2014)}]{Contreras:2013qua}
\bibinfo{author}{\bibfnamefont{E.}~\bibnamefont{Contreras}} \bibnamefont{and}
  \bibinfo{author}{\bibfnamefont{L.}~\bibnamefont{Leal}},
  \bibinfo{journal}{Int. J. Mod. Phys.} \textbf{\bibinfo{volume}{D23}},
  \bibinfo{pages}{1450047} (\bibinfo{year}{2014}), \eprint{1304.0778}.

\end{thebibliography}

\end{document}